# Electrical Parameters for Planar Transport in Graphene and 2-D Materials

Gananath Dash

Electron Devices Group, School of Physics, Sambalpur University, Jyoti Vihar, Burla, Sambalpur – 768019 (India)

### Abstract

Classical electrodynamics has been revisited with a view to recast the electrical parameters for planar transport in 2-dimensional (2-D) materials like graphene. In this attempt a new line integral, named "transverse line integral," with extensive applications in 2-D, is defined. Since the existing divergence theorem in not applicable in 2-D, we introduced a new divergence theorem. A new definition for the in-plane flux of any 2-D vector is introduced. A new vector named "electric vector potential" is defined and Gauss law is modified in terms of the 2-D flux of the new vector. The new Gauss law in presence of dielectric is obtained and a new "electric displacement vector" is defined for the 2-D materials. The conduction and displacement current densities in 2-D are defined. Resistance and resistivity in 2-D materials are discussed. The continuity equation for planar transport is derived.

# 1. Introduction

Electronic devices constitute a significant wing of the electrical engineering. Classical electrodynamics forms an important basis for proposing, designing and analysing the devices. But it goes without saying that, the laws of electrodynamics have been developed keeping in mind the 3-D (3-dimensional) nature of the physical world. In fact the material world is perceived with three physical dimensions customarily called length, breadth (or width), and depth (or thickness, also termed height). The appearance of Graphene in 2004 changed this perception; that's because graphene is a 2-D material [1]. Following graphene, we have now a host of materials, which are 2-D in nature. These include the hexagonal boron nitride (hBN) [2] and the transition metal di-chalcogenides $MX_2$ (M = Mo, Nb, W, Ta; X = S, Se, Te) [3]. Researchers believe that there are thousands of 2-D materials waiting to be discovered [4]. These materials have not only found extensive applications in the existing devices but also they have become instrumental in designing and fabricating quite a lot of new devices [5]. The question now arises whether all the laws of the 3-D electrodynamics will straight work for the new class of 2-D materials. The answer must be negative apparently because these materials have no thickness and hence no volume leading to the fact that quantities like volume integral, volume charge density and many more terms of the 3-D electrodynamics are indefinable. Therefore, it has become imperative to redefine them and develop a new electrodynamics for the 2-D materials by appropriately modifying the existing electrodynamics. One such attempt by Richard Lapidus [6] which originated much before the advent of graphene, considers a straightforward extension of the mathematics of 3+1 (3 space + 1 time) dimensional electrodynamics to 2+1 dimension. In essence he has developed the theory for a 2-D universe. As a result of the mathematical compulsion some strange outcome like $(1/r)$ dependence of Coulomb's law has evolved. But graphene and other 2-D materials cannot be regarded as constituting a 2-D universe as evident from the fact that, although the carriers in these materials are confined to 2-D, the field lines spread out into the 3-D space surrounding the material [7]. As a result, the 3-D coulomb's law is still valid in graphene except under certain restricted conditions [7]. In addition, there is no experimental evidence of inverse linear Coulomb's law holding good in graphene and other 2-D materials. We have therefore followed an alternative approach in this work to redefine the electrical parameters within the framework of 3-D Coulomb's law, with a view to figure out transport in 2-D materials. The paper is organised in the following way. The necessary mathematical background is developed in Sec. 2, Electrostatics of 2-D materials are discussed in Sec. 3, electric current density and continuity equation for 2-D are presented in Sec. 4 and finally the paper is concluded in Sec. 5.

## 2. Mathematical Background

2-D materials are devoid of the third dimension. In other words the depth or thickness of the material is so small that it is assumed to be zero. So we are left with only two dimensions, say the length and the breadth for which we may also designate it as a planar material. Customarily, we shall assume the material to be in the x-y plane of a Cartesian coordinate system. Since the z-dimension of the material is zero, carriers which will overcome the work function barrier will soon be out of the 2-D material. Hence, carrier motion inside the material cannot be envisaged in this direction. Therefore, for transport purpose the electrical quantities are defined in the x-y plane only. These properties will lead to the following modification in the calculus applicable to the 2-D materials.

### 2.1 Differential Calculus

Gradient of a scalar function $\phi$ is defined as

$$\vec{\nabla}_{2D}\phi = \hat{\imath}\frac{\partial \phi}{\partial x} + \hat{\jmath}\frac{\partial \phi}{\partial y}, \tag{1}$$

where $\hat{\imath}$ and $\hat{\jmath}$ are unit vectors in the x- and y-directions respectively. Similarly, the divergence of a vector function $\vec{F}$ is defined as

$$\vec{\nabla}_{2D}\cdot\vec{F} = \frac{\partial F_x}{\partial x} + \frac{\partial F_y}{\partial y}, \tag{2}$$

where $F_x$ and $F_y$ are the respective components of $\vec{F}$. Finally the curl of a vector function $\vec{F}$ is defined as

$$\vec{\nabla}_{2D}\times\vec{F} = \left(\hat{\imath}\frac{\partial}{\partial x} + \hat{\jmath}\frac{\partial}{\partial y}\right)\times\left(\hat{\imath}F_x + \hat{\jmath}F_y\right) = \hat{k}\left(\frac{\partial F_y}{\partial x} - \frac{\partial F_x}{\partial y}\right), \tag{3}$$

where $\hat{k}$ is a unit vector in the z-direction.

**2.2 Integral Calculus**

In integral calculus, we shall be concerned with only the line integral and the surface integral. Since 2-D material has no thickness, its volume cannot be defined. Consequently, the volume integral (of a 2-D material) does not exist.

The line integral, of a vector function $\vec{F}$ along a curve C from position *a* to *b* in the x-y plane is defined as

$$I_t = \int_C \vec{F}\cdot\hat{n}_\parallel \, dl, \tag{4}$$

where $\hat{n}_\parallel$ is a unit vector tangential to the curve C at the point under consideration (see Fig. 1(a)); the integration proceeds from point *a* to *b* covering the entire curve C. This is same as that defined for the 3-D material.

We now define a new line integral, not encountered in the study of 3-D electrodynamics. The conventional line integral defined by Eq. (4) has been termed as the *tangential* line integral by Charlie Harper [8]. This is indicative of the fact that there exists a scope to define a *transverse* line integral. But such a line integral had no application in 3-D electrodynamics. As a result of which it has not been considered so far. We will later see that this integral has a lot of applications in 2-D electrodynamics. We define the transverse line integral of a vector function $\vec{F}$ as

$$I_\perp = \int_C \vec{F}.\hat{n}_\perp \, dl, \tag{5}$$

where $\hat{n}_\perp$ is a unit vector normal to the tangent (to the curve C) at the point of consideration (see Fig.1(b)). As in the case of Eq. (4) the integration proceeds from point *a* to *b* covering the entire curve *C*.

The definition of surface integral is the same as that in 3-D and it can be defined for both a scalar function ($\phi$) and a vector function ($\vec{F}$). Considering a surface *S* in the x-y plane, these integrals are defined respectively as

$$I_s = \int_S \phi \, ds, \tag{6}$$

and

$$I_s = \int_S \vec{F}.\hat{n}_s \, ds, \tag{7}$$

where $\hat{n}_s$ is a unit vector perpendicular to the surface at the point under consideration; The integration is carried out over all elementary surfaces $ds$ covering the whole surface *S*. As stated earlier volume integral is indefinable in 2-D since a 2-D material has no volume.

**2.3 Theorems of Integral Calculus**

(A) *The gradient theorem*

The gradient theorem is the same as in 3-D i.e if a vector function $\vec{F}$ can be expressed as the gradient of a scalar function $\phi$ such as $\vec{F} = \vec{\nabla}_{2D}\phi$, then the tangential line integral of $\vec{F}$ does not depend upon the shape of the curve *C* (i.e. the path of integration) and reduces to the function of the end points *a* and *b* of the curve *C*. Mathematically

$$\int_C \vec{F}.\hat{n}_\parallel \, dl = \int_a^b \vec{\nabla}_{2D}\phi.\hat{n}_\parallel \, dl = \phi(b) - \phi(a). \tag{7}$$

(B) *The divergence theorem*

The divergence theorem for the 2-D case is completely different from that of the 3-D; that is because a volume integral is indefinable in the former. So we frame a new divergence theorem applicable to the 2-D materials which can be stated as "The surface integral of the divergence of a vector function $\vec{F}$ taken over any surface $S$ in the material plane, is equal to the transverse line integral of $\vec{F}$ taken over the closed curve $C$ constituting the boundary of $S$." We defer the proof of this theorem to Appendix – A and simply write the mathematical statement here:

$$\int_S \vec{\nabla}_{2D}.\vec{F}\, ds = \oint_C \vec{F}.\hat{n}_\perp dl. \tag{8}$$

(C) *The curl theorem*

It is the same as in 3-D case which may be stated for the 2-D case as follows: "The surface integral of the curl of a vector function $\vec{F}$ taken over any surface $S$ in the material plane, is the tangential line integral of $\vec{F}$ taken over the closed curve $C$ constituting the boundary of $S$." Mathematically

$$\int_S (\vec{\nabla}_{2D} \times \vec{F}).\hat{n}_s\, ds = \oint_C \vec{F}.\hat{n}_\parallel dl. \tag{9}$$

## 3. Electrostatics in 2-D

**3.1 Electric Field**

In order to investigate the electrical transport parameters in graphene we consider a general device structure as shown in Fig. 2. Since graphene is considered to be in the x-y plane the gate voltages (both top gate and back gate voltages) produce an electric field in the z-direction (we shall designate it as $\hat{k}E_z$). This electric field is not responsible for any charge transport since carrier motion is prevented in the z-direction due to confinement effect; on the other hand it produces an important effect in graphene called "the electrostatic doping."

Interested readers may go through the published literatures for more details [9]. It turns out that we are not concerned with $E_z$ in this study. Nonetheless, the drain to source bias produces an electric field in the graphene plane. We are interested in this electric field here. Although the composite effect of the gate and drain biases have the resultant electric field in some arbitrary direction, we shall be interested in the component of the resultant field in the x-y plane for transport purpose. An electric field in the material plane is a 2-D vector and can be written as

$$\vec{E} = \hat{\imath} E_x + \hat{\jmath} E_y .  \tag{10}$$

## 3.2 Flux of a Vector in 2-D

In a 2-D material, a surface perpendicular to the electric field $\vec{E}$ does not exist. Hence, the conventional definition of electric flux (used in 3-D) cannot be applied. Instead, we seek for a fresh definition of the 2-D electric flux. In doing so, we should see that the new definition maintains compatibility with the existing laws (in 3-D). Intuitively, we define it as the transverse line integral of the electric field vector along a closed curve $C$ in the material plane. Thus the electric flux in 2-D $\phi_{E2D}$, can be written as

$$\phi_{E2D} = \oint_C \vec{E} . \hat{n}_\perp dl .  \tag{11}$$

In fact, the flux of any 2-D vector through a closed curve $C$ may be defined as the transverse line integral of the vector along $C$.

## 3.3 Gauss Law in 2-D Free Space

Since Eq. (11) is the only way to define an electric flux for the electric field at Eq. (10), the question now arises whether the Gauss law still holds in 2-D. In 3-D it states that the "total electric flux through a closed surface is $1/\epsilon_0$ times the total charge enclosed by the surface." With the definition of electric flux as in Eq. (11), let us evaluate the flux through a circle of radius $r$ with a charge $q$ at the centre (Fig. 3),

$$\oint_C \vec{E}.\hat{n}_\perp dl = \oint_C \frac{1}{4\pi\epsilon_0}\frac{q}{r^2} r d\theta = \frac{q}{4\pi\epsilon_0 r}\int_0^{2\pi} d\theta = \frac{2\pi q}{4\pi\epsilon_0 r} = \frac{q}{2\epsilon_0 r}. \quad (12)$$

According to 3-D Gauss law, the result should have been $q/\epsilon_0$. Thus it is apparent that the 3-D Gauss law is not valid in 2-D specifically because of the inverse square Coulomb's law. The factor of 2 in the denominator could be conceded but the $1/r$ dependence of the flux would land it nowhere in the domain of a law. The way out for a 2-D Gauss law is to define a new vector whose flux evaluated as per Eq. (11) would do away with the $1/r$ dependence. Intuitively, electric potential is such a quantity, but it is a scalar. Just as the elementary area vector is defined (note that area is otherwise a scalar) for the purpose of surface integral, we define a new vector called the "electric vector potential" with magnitude equal to the potential $V$, and direction same as that of the electric field $\vec{E}$. The electric vector potential can thus be expressed as

$$\vec{A} = V\hat{n}_E, \quad (13)$$

where $\hat{n}_E$ is a unit vector in the direction of the electric field $\vec{E}$. We repeat the process at Eq. (12) for the electric vector potential $\vec{A}$ and obtain the following

$$\oint_C \vec{A}.\hat{n}_\perp dl = \oint_C \frac{1}{4\pi\epsilon_0}\frac{q}{r} r d\theta = \frac{q}{4\pi\epsilon_0}\int_0^{2\pi} d\theta = \frac{2\pi q}{4\pi\epsilon_0} = \frac{q}{2\epsilon_0}. \quad (14)$$

Incidentally we have this result for a circle as the closed curve C. But it is not difficult to visualise that the result will be the same for any closed curve C enclosing the charge $q$, since the result is independent of the distance of the curve from the charge. For the same reason, and the principle of superposition, the result will also be true for multiple number of charges or a given charge distribution enclosed by the curve C. Eventually we have arrived at the 2-D Gauss law which can be formally stated as follows. "The total flux of the electric vector potential through any closed curve is $1/2\epsilon_0$ times the total charge enclosed by the curve." Mathematically

$$\oint_C \vec{A}.\hat{n}_\perp dl = \frac{Q_{enc}}{2\epsilon_0}, \quad (15)$$

where $Q_{enc}$ is the total charge enclosed by the curve.

We now seek for a differential form of the Gauss law. Making use of the divergence theorem from Eq. (8), Gauss law at Eq. (15) can be converted to the following form

$$\int_S \vec{\nabla}_{2D}.\vec{A}\, ds = \frac{Q_{enc}}{2\epsilon_0}. \quad (16)$$

It may be stressed upon here that a 2-D material does not have "volume charge density." The role of this quantity is played by the "surface charge density" (customarily called the "sheet charge density") in 2-D case which is defined as the charge per unit area of the 2-D space. If $\rho_s$ be the sheet charge density, then one can write

$$Q_{enc} = \int_S \rho_s\, ds. \quad (17)$$

Using Eq. (16) and (17) we get

$$\int_S \vec{\nabla}_{2D}.\vec{A}\, ds = \frac{1}{2\epsilon_0} \int_S \rho_s\, ds. \quad (18)$$

Since the surface $S$ is arbitrary, Eq. (18) implies the equality of the integrands leading to

$$\vec{\nabla}_{2D}.\vec{A} = \frac{\rho_s}{2\epsilon_0}. \quad (19)$$

This is the differential form of Gauss law in 2-D. Two points of difference from the 3-D case are worth noting in this equation: (i) the electric potential vector takes the place of the electric field, and (ii) the volume charge density is replaced with the sheet charge density.

### 3.4 Gauss Law in Dielectric

(A) *Polarisation*

In most of the cases the 2-D material is a dielectric. Hence the Gauss law should take care of the dielectric effect. When the electric field $\vec{E}$ passes through the dielectric of the 2-D material, it generates numerous tiny dipoles each with a dipole moment. The collective dipole moment of all the dipoles in unit area of the 2-D material should now be defined as the polarisation vector $\vec{P}_{2D}$ with components in the material plane only (note the difference from the 3-D case where the polarisation vector is defined as dipole moment per unit volume with 3 components). The polarisation vector $\vec{P}_{2D}$ has the same direction as that of the electric field $\vec{E}$. In essence, the electric field, as in 3-D case, displaces the centres of positive and negative charge of the atoms giving rise to a distribution of new charge called "the bound charge or the polarisation charge." The Gauss law in dielectric, now boils down to inclusion of the new charge. Denoting the bound charge density per unit area as $\rho_{sb}$, the Gauss law at Eq. (19) gets modified to

$$\vec{\nabla}_{2D}.\vec{A} = \frac{1}{2\epsilon_0}(\rho_s + \rho_{sb}). \tag{20}$$

In order to distinguish from the bound charge, the charge density $\rho_s$ is customarily termed as the free charge density (we have avoided an additional subscript 'f' for the free charge density to make the notation simplified). It will be our endeavour now to get an expression for $\rho_{sb}$.

(B) *Bound Charge Density*

Consider a rectangular section PQRS of length $l$ (parallel to the electric field) and width $w$ of the 2-D material as shown in Fig. 4(a). As a result of polarisation, bound charges $+q$ and $-q$ appear across the edges QR and PS respectively (in general, for an arbitrary field direction, bound charges should also appear across PQ and RS, but for simplicity we have chosen the length of the rectangular section parallel to the electric field leading to no bound charge across PQ and RS). The dipole moment of the rectangular section is $ql$. Using the definition of 2-D polarisation (dipole moment per unit area) we can write

$$ql = |\vec{P}_{2D}|lw \quad \Rightarrow \quad |\vec{P}_{2D}| = \frac{q}{w} = \rho_l \quad \Rightarrow \quad \rho_l = \vec{P}_{2D}.\hat{n}_{\perp QR}, \tag{21}$$

where $\hat{n}_{\perp QR}$ is a unit vector perpendicular to QR directed outward. Note that the quantity $\rho_l$ has the dimension of charge per unit line length and hence it is defined as the "line charge density." Now to generalise the concept to a section of the 2-D material enclosed by a curve $C$ of arbitrary shape refer to Fig. 4(b). Divide the curve into small line elements $dl$. The line charge density $\rho_l$ in each line element is $\rho_l = \vec{P}_{2D} \cdot \hat{n}_\perp$ where as usual $\hat{n}_\perp$ a unit vector normal to the line element $dl$. Finally, the total bound charge $Q_b$ across the curve is obtained by integrating $\rho_l dl$ over the whole curve $C$,

$$Q_b = \oint_C \rho_l dl = \oint_C \vec{P}_{2D} \cdot \hat{n}_\perp dl . \tag{22}$$

The bound charge across a closed curve $C$ in a 2-D material is thus equal to the transverse line integral of the polarisation vector. The bound charge $Q_b$ appearing across the boundary line $C$ is the result of an equal and opposite amount of charge being displaced from within the surface $S$ enclosed by $C$. Consequently the bound surface charge equals $-Q_b$. As a result, we can write

$$\int_S \rho_{sb} ds = -Q_b = -\oint_C \vec{P}_{2D} \cdot \hat{n}_\perp dl . \tag{23}$$

Using the divergence theorem in 2-D from Eq. (8), the above equation is converted to

$$\int_S \rho_{sb} ds = -\int_S \vec{\nabla}_{2D} \cdot \vec{P}_{2D} \, ds . \tag{24}$$

As in other cases, since this equation holds for any surface $S$, the integrands must be equal. Thus

$$\rho_{sb} = -\vec{\nabla}_{2D} \cdot \vec{P}_{2D} . \tag{25}$$

(C) *Gauss law*

Now we return to Eq. (20) and substitute Eq. (25) into it to get

$$\vec{\nabla}_{2D}\cdot\vec{A} = \frac{1}{2\epsilon_0}\left(\rho_s - \vec{\nabla}_{2D}\cdot\vec{P}_{2D}\right), \tag{26}$$

which on rearranging gives

$$\vec{\nabla}_{2D}\cdot\left(2\epsilon_0\vec{A} + \vec{P}_{2D}\right) = \rho_s. \tag{27}$$

This is 2-D Gauss law in dielectric. The quantity under parenthesis in Eq. (27) is defined as the electric displacement vector $\vec{D}_{2D}$, a 2-D vector in this case:

$$\vec{D}_{2D} = 2\epsilon_0\vec{A} + \vec{P}_{2D}. \tag{28}$$

This expression drastically differs from the 3-D case. With the defining of the electric displacement vector, we have another differential form of Gauss law in 2-D dielectric such as

$$\vec{\nabla}_{2D}\cdot\vec{D}_{2D} = \rho_s. \tag{29}$$

The total charge in free space enclosed by any curve $C$ can be determined using Gauss law at Eq. (15). Now to calculate the same in a 2-D dielectric, we start with Eq. (29) and integrate it over the surface $S$ enclosed by $C$,

$$\int_S \vec{\nabla}_{2D}\cdot\vec{D}_{2D}\, ds = \int_S \rho_s\, ds. \tag{30}$$

Using the divergence theorem in 2-D, Eq. (30) can be written as

$$\int_S \vec{D}_{2D}\cdot\hat{n}_\perp\, dl = Q_{enc}. \tag{31}$$

The total charge within a surface S enclosed by the curve C in a 2-D dielectric can thus be determined from the transverse line integral of the electric displacement vector along C.

(D) *Evaluation of* $\vec{D}_{2D}$

We assume a linear dielectric. In 3-D case this assumption leads to $\vec{P} = \epsilon_0\chi_e\vec{E}$ where $\chi_e$ is a proportionality constant named electric susceptibility. For a given device geometry, it is reasonable to assume that the electric field monotonically increases with the applied voltage. While the actual dependence of polarisation vector on the potential is fairly complicated, we use the same analogy as done in the 3-D case and assume $\vec{P}_{2D} = \epsilon_0 \chi_e \vec{A}$. Then from Eq. (28) we get

$$\vec{D}_{2D} = 2\epsilon_0 \vec{A} + \epsilon_0 \chi_e \vec{A} = \epsilon_0 (2 + \chi_e)\vec{A} = \epsilon_0 (1 + K)\vec{A}, \tag{32}$$

where $K = 1 + \chi_e$ is defined as the dielectric constant of the material. Now the 2-D permittivity of the material has the expression

$$\epsilon_{2D} = \epsilon_0 (1 + K), \tag{33}$$

and then we can express the electric displacement vector in 2-D as

$$\vec{D}_{2D} = \epsilon_{2D} \vec{A}. \tag{34}$$

This relation substantially differs from the 3-D case.

## 4. Electric Current Density in 2-D

**4.1 Conduction Current Density**

The flow of charge in a 2-D material is restricted to the x-y plane. Electric current density in 3-D is defined as the current per unit surface area passing normal to the surface. But in 2-D case no surface is available normal to the current which is constrained to flow in the material plane. It turns out that the 3-D definition of current density cannot be applied to the 2-D material. Therefore, an alternative definition is indispensable. We use the same idea as in defining the electric flux for a fresh definition of the 2-D current density. Specifically we define it as a 2-D vector with magnitude "current per unit line length passing normal to the line" and direction that of the "current flow." For obvious reason we shall designate the current density as "line current density" in 2-D case and "surface current density" in 3-D. Considering a simple geometry in a 2-D material as in Fig. 5(a) the conduction current density (magnitude) will be given by

$$J_{c2D} = \frac{I_c}{W} \quad \Rightarrow \quad I_c = J_{c2D}W = \vec{J}_{c2D} \cdot \hat{n}_{\perp W} W, \tag{35}$$

where $\hat{n}_{\perp W}$ is a unit vector normal to the width $W$ directed outward. To determine the current through an arbitrary line specified by a curve $C$ as in Fig. 5(b), we divide the curve into line elements $dl$ and determine the current element in each as

$$dI_c = \vec{J}_{c2D} \cdot \hat{n}_{\perp} dl. \tag{36}$$

The total current passing through the line denoted by $C$ is obtained by integration of Eq.(36),

$$I_c = \int_C dI_c = \int_C \vec{J}_{c2D} \cdot \hat{n}_{\perp} dl. \tag{37}$$

In other words, the current passing through any line in a 2-D material is the transverse line integral of the current density along that line.

### 4.2 Displacement Current Density

For time-varying electric field, we have an additional component of the current density termed as the displacement current density as suggested by Maxwell. For a 2-D material the line current density for the displacement current can be defined as

$$\vec{J}_{d2D} = \frac{\partial \vec{D}_{2D}}{\partial t}, \tag{38}$$

where $\vec{D}_{2D}$ is the electric displacement vector in the 2-D material. Adopting the same procedure as for the conduction current, the displacement current can be obtained as

$$I_d = \int_C \vec{J}_{d2D} \cdot \hat{n}_{\perp} dl. \tag{39}$$

The total current density is the sum of the conduction and displacement current densities,

$$\vec{J}_{2D} = \vec{J}_{c2D} + \vec{J}_{d2D}. \tag{40}$$

### 4.3 Resistance and resistivity

Resistance ($R$) in 2-D material has the same definition as in 3-D case: $V = IR$, where $V$ is the potential difference and $I = I_c$. However, resistivity $\rho_R$ is different. In 3-D it is defined as the resistance of unit length (parallel to the direction of current flow) and unit area of cross section (normal to the direction of current flow) of the material. But in 2-D case, for the planar conduction, area normal to the direction of current flow is unavailable. For a way out, consider a rectangular section of the material as in Fig. 6. The electric field in the material is $\vec{E} = \hat{\imath}\, V/L$. We invoke the famous Ohm's law and get

$$\vec{J}_{c2D} = \sigma \vec{E} \quad \Rightarrow \quad \vec{E} = \rho_R \vec{J}_{c2D}, \tag{41}$$

where $\sigma$ is the conductivity and $\rho_R = 1/\sigma$ is the resistivity of the material. If we use the definition of current density at Eq. (35) and the expression for the electric field above, we get

$$\frac{V}{L}\hat{\imath} = \rho_R \frac{I_c}{W}\hat{n}_{\perp W} \quad \Rightarrow \quad R = \rho_R \frac{L}{W}. \tag{42}$$

We arrived at a new law of resistance for the 2-D materials: "The resistance of a 2-D material is directly proportional to the length (parallel to current flow) and inversely proportional to the width (normal to current flow)." The proportionality constant is defined as the resistivity of the material. Accordingly,

$$\rho_R = R\,\frac{W}{L}. \tag{43}$$

In other word, the resistivity of a 2-D material is the resistance of unit length and unit width of the material.

### 4.4 Continuity equation

Consider a closed curve $C$ in the plane of the 2-D material. The current passing out of this curve is determined by the then transverse line integral of the current density along the curve. This is also the rate of charge flowing out of the area enclosed by $C$ which in turn is equal to the rate of decrease of charge enclosed by C. All of these can be put together into the following equation

$$I = \int_C \vec{J}_{c2D} \cdot \hat{n}_\perp dl = \frac{dQ_{out}}{dt} = -\frac{dQ_{enc}}{dt}. \qquad (44)$$

Using Eq. (17) for the enclosed charge, Eq. (44) becomes

$$\int_C \vec{J}_{c2D} \cdot \hat{n}_\perp dl = -\frac{\partial}{\partial t}\int_S \rho_s ds, \qquad (45)$$

where S is the surface area enclosed by C. Now using the divergence theorem in 2-D, Eq. (45) can be converted to

$$\int_S \vec{\nabla}_{2D} \cdot \vec{J}_{c2D} ds = -\frac{\partial}{\partial t}\int_C \rho_s ds. \qquad (46)$$

Since the curve C, and for that matter the surface area S, is arbitrary it implies that the integrands of Eq. (46) must be equal. Rearranging we get

$$\frac{\partial \rho_s}{\partial t} = -\vec{\nabla}_{2D} \cdot \vec{J}_{c2D} \quad \Rightarrow \quad \frac{\partial \rho_s}{\partial t} + \vec{\nabla}_{2D} \cdot \vec{J}_{c2D} = 0. \qquad (47)$$

This is the continuity equation in 2-D. Note that the difference in definition of $\vec{J}_{c2D}$ from the 3-D case is essential to cope with the surface charge density (instead of volume charge density) involved in the 2-D case.

## 5. Conclusion

Electrical transport parameters for in-plane conduction in 2-D materials have been obtained by revisiting the classical 3-D electrodynamics within the framework of inverse square Coulomb's law. For this purpose gradient, divergence and curl suitable for 2-D have been first defined. A new line integral, with extensive applications in 2-D, named "transverse line integral" (in addition to the existing one which we rename as "tangential line integral") has been defined. Since the existing divergence theorem in not applicable in 2-D, a new divergence theorem has been introduced. The roles of in-plane and cross-plane electric fields in the 2-D material have been separated. As the conventional electric flux for the planar field is indefinable, a new definition for the in-plane electric flux has been introduced. Gauss law is tested for the defined planar electric flux and is found to be invalid. So a new vector named "electric vector potential" has been defined. Gauss law has been modified in terms of the 2-D flux of the new vector. Differential form of the new Gauss law has been derived. The effect of dielectric on the new Gauss law has been discussed. Expression for the bound charge density per unit area has been obtained in terms of the polarisation vector. The new Gauss law in presence of dielectric has been obtained and a new "electric displacement vector" has been defined for the 2-D materials. The definitions of conduction and displacement current densities have been modified. Resistance and resistivity in 2-D materials have been discussed and continuity equation for 2-D has been derived.

**References**

[1] K. S. Novoselov *et al.*, "Electric field effect in atomically thin carbon films," *Science*, vol. 306, no. 5696, pp. 666–669, 2004.

[2] D. Pacilé, J. C. Meyer, Ç. Ö. Girit, and A. Zettl, "The two dimensional phase of boron nitride: Few-atomic-layer sheets and suspended membranes," *Appl. Phys. Lett.*, vol. 92, p. 133107, Apr. 2008.

[3] Y. Ding *et al.*, "First principles study of structural, vibrational and electronic properties of graphene-like $MX_2$ (M = MO, Nb, W, Ta; X = S, Se, Te) monolayers," *Phys. B*, vol. 406, no. 11, pp. 2254–2260, 2011.

[4] Nicolas Mounet *et al.*, "Two-dimensional materials from high-throughput computational exfoliation of experimentally known compounds", *Nature Nanotechnology*, vol. 13, pp. 246–252, 2018.

[5] G. N. Dash, Satya R. Pattanaik, and Sriyanka Behera, "Graphene for electron devices: the panorama of a decade," *Journal of the Electron Devices Society*, vol. 2, no. 5, pp. 77-104, 2014.

[6] I. Richard Lapidus, "Classical electrodynamics in a universe with two space dimensions," *American Journal of Physics*, vol. 50, pp. 155, 1982.

[7] Valeri N. Kotov, Bruno Uchoa, and A. H. Castro Neto, "1/N expansion in correlated graphene," *Phys. Rev. B*, vol 80, pp. 165424, 2009.

[8] Charlie Harper, "Introduction to Mathematical Physics," PHI Learning Pvt. Ltd., New Delhi, 2009.

[9] Gaurav Gupta, Bijoy Rajasekharan, and Raymond J. E. Hueting, "Electrostatic Doping in Semiconductor Devices," *IEEE Transactions on Electron Devices*, vol. 64, no. 8, pp. 3044 – 3055, 2017.

**Appendix – A: Proof of Divergence Theorem in 2-D**

Consider a surface S in the x-y plane bounded by the rectangle "abcda" of length L and width W as show in Fig. 7. For any vector function $\vec{F}(x,y)$ we have to prove that

$$\int_S \vec{\nabla}_{2D}.\vec{F}(x,y)\, ds = \oint_{abcda} \vec{F}(x,y).\hat{n}_\perp dl. \tag{A1}$$

Evaluate the LHS over the surface S

$$\int_S \vec{\nabla}_{2D}.\vec{F}(x,y)\, ds$$

$$= \int_0^W \int_0^L \left(\frac{\partial F_x(x,y)}{\partial x} + \frac{\partial F_y(x,y)}{\partial y}\right) dx\, dy$$

$$= \int_0^W \left(\int_0^L \frac{\partial F_x(x,y)}{\partial x} dx\right) dy + \int_0^L \left(\int_0^W \frac{\partial F_y(x,y)}{\partial y} dy\right) dx \tag{A2}$$

Take the first term on the RHS and simplify

$$\int_0^W \left(\int_0^L \frac{\partial F_x(x,y)}{\partial x} dx\right) dy$$

$$= \int_0^W \left(\int_0^L dF_x(x,y)\right) dy$$

$$= \int_0^W [F_x(L,y) - F_x(0,y)]\, dy$$

$$= \int_0^W F_x(L,y)\, dy - \int_0^W F_x(0,y)\, dy \tag{A3}$$

With reference to Fig. 7 note that $F_x(L,y)$ refers to the value of $F_x$ at $x = L$ and hence the y integration in the first tem of Eq. (A3) should proceed from *b* to *c*. Similarly the second integration which refers to the value of $F_x$ at $x = 0$ should proceed from *a* to *d*. So the first term of Eq. (A2) reduces to

$$\int_b^c F_x(L, y)\, dy - \int_a^d F_x(0, y)\, dy \tag{A4}$$

To relate these integrations to line integrals we tabulate the line elements and unit vectors for the four arms of the rectangle in Table A1. The convention used to obtain $\hat{n}_\perp$ from $\hat{n}_\parallel$ is by rotating clockwise through $\pi/2$.

Table A1: Unit vectors and line elements for the four sides of the rectangle at Fig. 7

| Sides of rectangle Parameters | ab | bc | dc | ad |
|---|---|---|---|---|
| $\hat{n}_\parallel$ | $\hat{\imath}$ | $\hat{\jmath}$ | $\hat{\imath}$ | $\hat{\jmath}$ |
| $\hat{n}_\perp$ | $-\hat{\jmath}$ | $\hat{\imath}$ | $-\hat{\jmath}$ | $\hat{\imath}$ |
| $dl$ | $dx$ | $dy$ | $dx$ | $dy$ |

Multiplying expression (A4) with $\hat{\imath}.\hat{\imath}$ (since $\hat{\imath}.\hat{\imath} = 1$, the same does not affect it) and making use of the values from Table (A1) we get

$$\int_b^c F_x(L, y)\, dy - \int_a^d F_x(0, y)\, dy$$

$$= \int_b^c F_x(L, y)\, (\hat{\imath}.\hat{\imath}) dy - \int_a^d F_x(0, y)\, (\hat{\imath}.\hat{\imath})\, dy$$

$$= \int_b^c \hat{\imath} F_x(L, y).\hat{\imath} dy - \int_a^d \hat{\imath} F_x(0, y).\hat{\imath} dy$$

$$= \int_b^c \hat{\imath} F_x(L, y).\hat{n}_\perp dl - \int_a^d \hat{\imath} F_x(0, y).\hat{n}_\perp dl$$

$$= \int_b^c \hat{\imath} F_x(L, y).\hat{n}_\perp dl + \int_d^a \hat{\imath} F_x(0, y).\hat{n}_\perp dl \tag{A5}$$

These are the transverse line integrals of $\hat{\imath} F_x$ along two sides of the rectangle, *bc* and *da*. Now considering the other two sides *ab* and *cd* we find that

$$\int_a^b \hat{i}F_x(x,0) \cdot \hat{n}_\perp dl = \int_a^b \hat{i}F_x(x,0) \cdot (-\hat{j})dx = 0 \qquad (A6)$$

and

$$\int_c^d \hat{i}F_x(x,W) \cdot \hat{n}_\perp dl = -\int_d^c \hat{i}F_x(x,W) \cdot (-\hat{j})dx = 0 \qquad (A7)$$

Making use of relations (A3) through (A7), we can write

$$\int_0^W \left( \int_0^L \frac{\partial F_x(x,y)}{\partial x} dx \right) dy$$

$$= \int_a^b \hat{i}F_x(x,0) \cdot \hat{n}_\perp dl + \int_b^c \hat{i}F_x(L,y) \cdot \hat{n}_\perp dl + \int_c^d \hat{i}F_x(x,W) \cdot \hat{n}_\perp dl + \int_d^a \hat{i}F_x(0,y) \cdot \hat{n}_\perp dl$$

$$= \oint_{abcda} \hat{i}F_x(x,y) \cdot \hat{n}_\perp dl \qquad (A8)$$

Following a similar procedure, it can be shown that

$$\int_0^L \left( \int_0^W \frac{\partial F_y(x,y)}{\partial y} dy \right) dx = \oint_{abcda} \hat{j}F_y(x,y) \cdot \hat{n}_\perp dl \qquad (A9)$$

Using Eqs. (A2), (A8) and (A9) we get

$$\int_S \vec{\nabla}_{2D} \cdot \vec{F}(x,y) \, ds$$

$$= \oint_{abcda} \hat{i}F_x(x,y) \cdot \hat{n}_\perp dl + \oint_{abcda} \hat{j}F_y(x,y) \cdot \hat{n}_\perp dl$$

$$= \oint_{abcda} \{\hat{i}F_x(x,y) + \hat{j}F_y(x,y)\} \cdot \hat{n}_\perp dl$$

$$= \oint_{abcda} \vec{F}(x,y) \cdot \hat{n}_\perp dl \qquad (A10)$$

This proves divergence theorem in 2-D.

**Figures:**

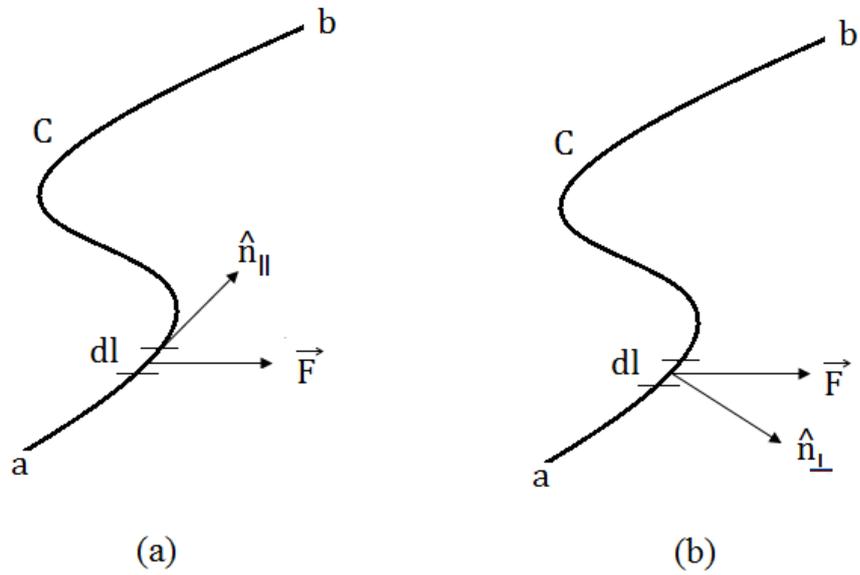

Fig. 1. (a) Tangential line integral, (b) Transverse line integral

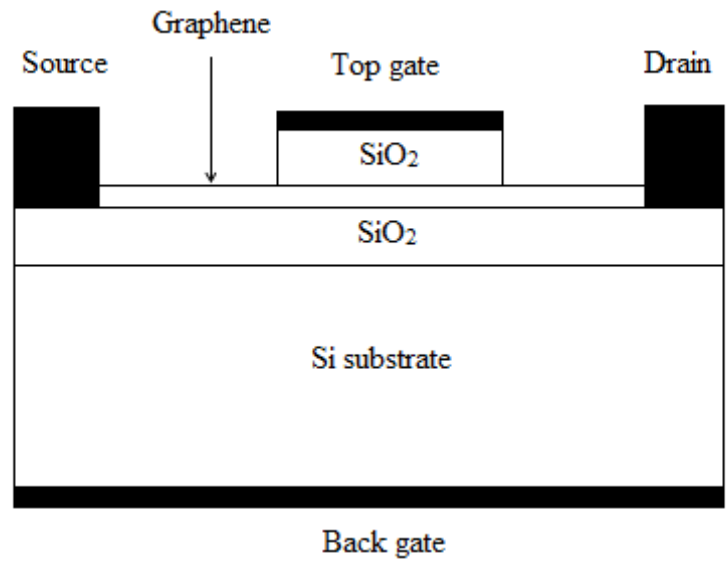

Fig. 2. Graphene as a planar 2-D material used in a typical device structure

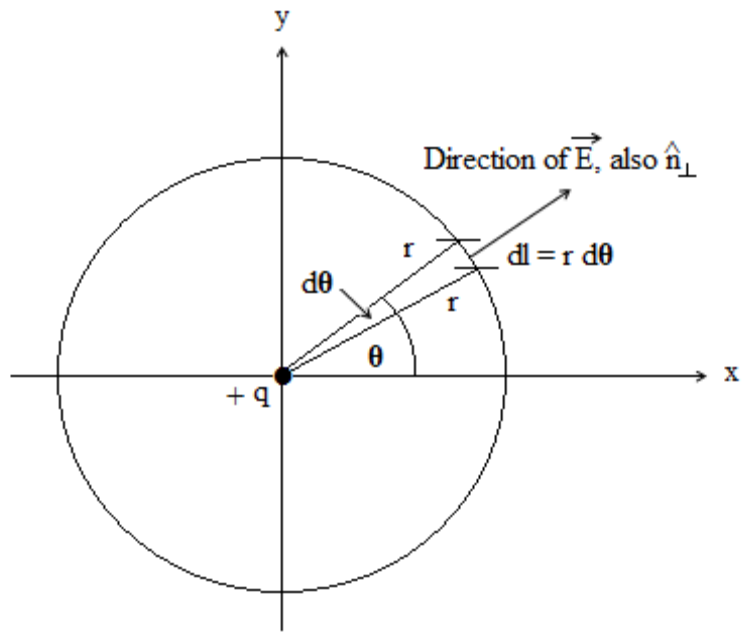

Fig. 3. Evaluation of 2-D electric flux through a circle of radius r with a point charge +q at the centre.

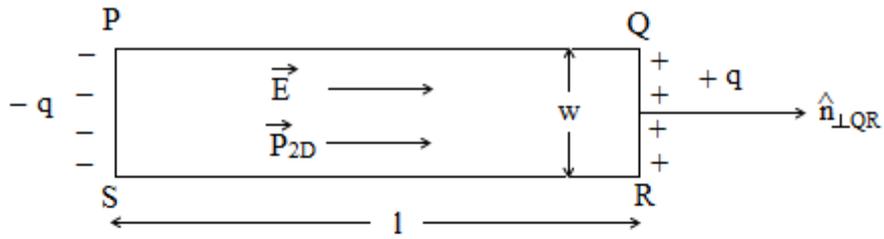

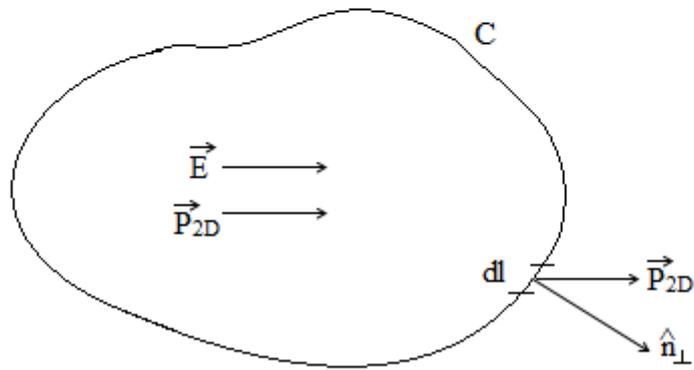

Fig. 4. To evaluate a relation between the 2-D polarisation vector and line charge density (a) for a rectangular section of the material, (b) for a section of the material with an arbitrary shape denoted by a closed curve C.

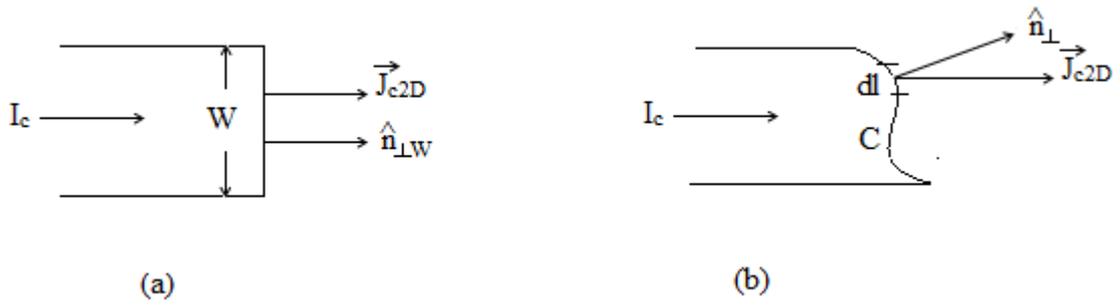

Fig. 5. Relation between current and 2-D current density, (a) considering a straight line section of width W, (b) considering a curved line section with arbitrary shape C.

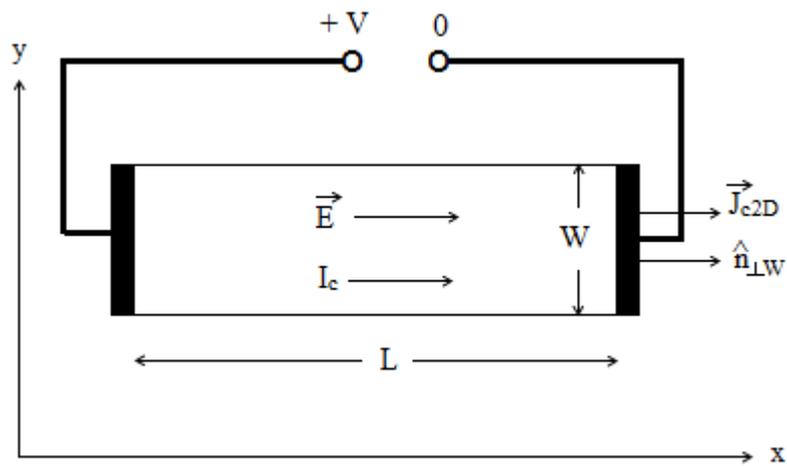

Fig. 6. A rectangular section of the 2-D material with length L (parallel to current flow direction) and width W (perpendicular to current flow direction) with an applied voltage V for the analysis of resistivity of the material.

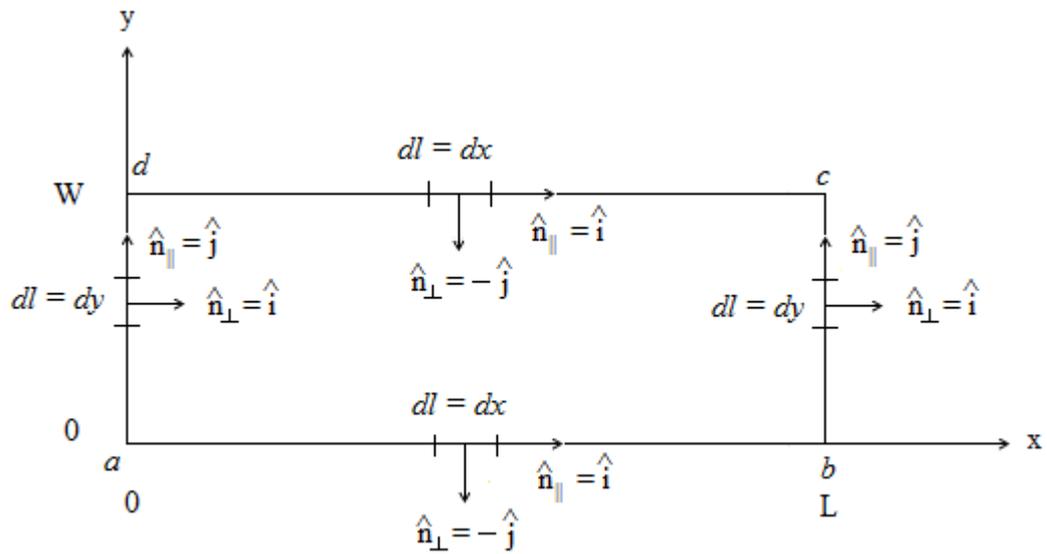

Fig. 7. A surface S bounded by a rectangle *abcd* of length L and width W considered for the proof of 2-D divergence theorem. The line elements along with the tangential and transverse unit vectors in different branches of the rectangle are shown.